\documentclass[journal]{IEEEtran}

\usepackage{graphics} % for pdf, bitmapped graphics files
\usepackage{epstopdf}
\usepackage{epsfig} % for postscript graphics files
\usepackage{subfig}
\usepackage{mathptmx} % assumes new font selection scheme installed
\usepackage{times} % assumes new font selection scheme installed
\usepackage{amsmath} % assumes amsmath package installed
\usepackage{amssymb}  % assumes amsmath package installed
\usepackage{algorithm}
\usepackage{algorithmic}
\usepackage{setspace}

\usepackage{setspace}
\usepackage{threeparttable}
\usepackage{multirow}
\usepackage{booktabs}
\usepackage{makecell}
\usepackage{float}
\usepackage{color}
\usepackage[colorlinks,linkcolor=red,anchorcolor=blue,citecolor=green]{hyperref}
\usepackage{bm}

\makeatletter

\newcommand{\Rmnum}[1]{\expandafter\@slowromancap\romannumeral #1@}

\newsavebox{\mybox}
\newcolumntype{X}[1]{>{\begin{lrbox}{\mybox}}l<{\end{lrbox}\makecell[#1]{{\usebox\mybox}}}}
\makeatother

\begin{document}

\title{A Grouping Based Cooperative Driving Strategy for CAVs Merging Problems}

\author{Huile~Xu,
        Shuo~Feng,
        Yi~Zhang,~\IEEEmembership{Member,~IEEE,}
        and~Li~Li,~\IEEEmembership{Fellow,~IEEE}% <-this % stops a space
\thanks{Manuscript received in \today; This work was supported in part by National Natural Science Foundation of China under Grant 61673233 and Beijing Municipal Science and Technology Committee Project under Grants D171100000317002 and ZC179074Z. (Corresponding author is \textit{Li Li}). }
\thanks{H. Xu is with Department of Automation, Tsinghua University, Beijing 100084, China. (Email: hl-xu16@mails.tsinghua.edu.cn)}% <-this % stops a space
\thanks{S. Feng is with Department of Automation, Tsinghua University, Beijing 100084, China. (Email: s-feng14@mails.tsinghua.edu.cn)}
\thanks{Y. Zhang is with Department of Automation, BNRist, Tsinghua University, Beijing 100084, China and also with the Tsinghua-Berkeley Shenzhen Institute (TBSI), Tower C2, Nanshan Intelligence Park 1001, Xueyuan Blvd., Nanshan District, Shenzhen 518055, China. (Email: zhyi@tsinghua.edu.cn)}% <-this % stops a space
\thanks{L. Li is with Department of Automation, BNRist, Tsinghua University, Beijing 100084, China. (Tel: +86(10)62782071, Email: li-li@tsinghua.edu.cn).}}

\maketitle

\begin{abstract}
In general, there are two kinds of cooperative driving strategies, planning based strategy and ad hoc negotiation based strategy, for connected and automated vehicles (CAVs) merging problems.
The planning based strategy aims to find the global optimal passing order, but it is time-consuming when the number of considered vehicles is large.
In contrast, the ad hoc negotiation based strategy runs fast, but it always finds a local optimal solution.
In this paper, we propose a grouping based cooperative driving strategy to make a good tradeoff between time consumption and coordination performance.
The key idea is to fix the passing orders for some vehicles whose inter-vehicle headways are small enough (e.g., smaller than the pre-selected grouping threshold).
From the viewpoint of optimization, this method reduces the size of the solution space.
A brief analysis shows that the sub-optimal passing order found by the grouping based strategy has a high probability to be close to the global optimal passing order, if the grouping threshold is appropriately chosen. A series of simulation experiments are carried out to validate that the proposed strategy can yield a satisfied coordination performance with less time consumption and is promising to be used in practice.
\end{abstract}


\IEEEpeerreviewmaketitle

\section{Introduction}

\IEEEPARstart{T}{raffic} congestion has caused huge loss to society and aroused wide concern in recent years\cite{schrank20152015,knoop2009microscopic}. Researchers had found that orderless merging at on-ramps is one of the main causes of traffic congestion and needs to be carefully handled \cite{milanes2011automated}.

The emergence of Connected and Automated Vehicles (CAVs) provides a promising way for solving merging problems. With the aid of vehicle-to-vehicle (V2V) and vehicle-to-infrastructure (V2I) communication, CAVs can obtain real-time operational data of adjacent vehicles and receive control actions\cite{li2014survey, korkmaz2010supporting, sukuvaara2009wireless}. It has become a common vision that CAVs will increasingly appear on the road in the near future and help to alleviate traffic congestion\cite{uno1999merging,li2017recasting}.

Along with the development of CAVs, researchers became interested in finding an efficient cooperative driving method for CAVs merging problems. It is pointed out in \cite{li2005cooperative} and \cite{li2006cooperative} that the key to the merging problem is to determine the optimal passing order. As summarized in \cite{li2014survey,meng2017analysis}, there are two kinds of cooperative driving strategies, planning based strategy and ad hoc negotiation based strategy, for determining the passing order.

\underline{\emph{Planning based strategy:}} The planning based strategy considers CAVs within a certain scope of the merging zone and provides long-term scheduled control actions for CAVs. It tries to enumerate all possible passing orders to find the global optimal solution. Most state-of-the-art studies transfer the merging problem into various optimization problems \cite{rios2016survey,rios2016automated,fayazi2017optimal,muller2016intersection,hult2016coordination}, such as mixed integer linear programming (MILP), receding horizon control (RHC). However, the time consumption for solving the problem increases sharply as the number of vehicle increases, which makes these methods difficult to be applied in practice.

\begin{table*}[htbp]
  \centering
  \caption{Nomenclature and representative values}
    \begin{tabular}{lll}
    \toprule
    Symbol & Meaning & Value\\
    \midrule
    \multicolumn{3}{l}{\emph{The symbols below are treated as constants}}\\
    $\Delta t_1$ &  The minimum safety gap between two CAVs on the same movement   & 1.5$s$  \\
    $\Delta t_2$ &  The minimum safety gap between two CAVs on the conflict movements   & 2$s$  \\
    $a_{max},a_{min}$ & The maximum and minimum acceleration  & -3$m/s^2$, 3$m/s^2$ \\
    $v_{max},v_{min}$ & The maximum and minimum velocity & 10$m/s$, 0$m/s$\\
    \midrule
    \multicolumn{3}{l}{\emph{The symbols below are treated as variables}}\\
    $\delta$     & Grouping threshold \\
    $x_i(t)$     & The location of CAV$_i$ at time $t$ \\
    $v_i(t)$     & The velocity of CAV$_i$ at time $t$\\
    $a_i(t)$     & The acceleration of CAV$_i$ at time $t$\\
    $t_{min,i}$  & The minimum access time of CAV$_i$\\
    $t_{assign,i}$ & The assigned access time of CAV$_i$\\
    \midrule
    \multicolumn{3}{l}{\emph{The symbols below are treated as functions or sets}}\\
    $J$ & Objective function \\
    $\bm{B}$ & Search space  \\
    $S$ & Selected set  \\
    $G$ & Good enough set  \\

    \bottomrule
    \end{tabular}%
  \label{tab:Mainnotation}%
\end{table*}%

\underline{\emph{Ad hoc negotiation based strategy:}} Ad hoc negotiation based strategy considers CAVs that are about to arrive at the merging zone and formulate short-term scheduled control actions via bilateral negotiations. It uses greedy search algorithms to determine the passing order and always lead the passing order to be roughly first-in-first-out (FIFO)\cite{dresner2004multiagent,dresner2008multiagent}. This strategy has a fast online implementation\cite{zhang2016optimal,de2017traffic}. However, they cannot guarantee that the passing order is global or good enough \cite{meng2017analysis,vasirani2012market,vasirani2009market}.

To overcome the above limitations, we propose a grouping based cooperative driving strategy for CAVs merging problems.
The key idea of grouping is to consider some vehicles (whose inter-vehicle headways are small enough) as a whole in planning.
This will narrow down the size of solution space and thus save planning time.
The idea of grouping had been initialized in \cite{li2005cooperative,li2006cooperative} for faster V2X communications.
Now, we apply the same trick for planning of cooperative driving.

First, CAVs within a control zone will be self-organized into several groups by a grouping method. Generally, if the headway between two consecutive CAVs is smaller than a grouping threshold, then they will be grouped into the same group. An adaptive grouping threshold is designed to control the size of groups. The maximum number of groups is fixed to limit the maximum time consumption.

Second, the passing order of CAVs in the same group is consecutively fixed. Therefore, the number of possible passing orders is largely reduced. Planning in such a reduced solution space will lead to a sub-optimal solution.

A brief analysis shows that the sub-optimal passing order found by the grouping based strategy has a high probability to be close to the global optimal passing order, if the grouping threshold is appropriately chosen.
To validate this finding, some simulation experiments are carried out.
Results indicate that the proposed strategy for merging problems can yield a good enough passing order with little time consumption.

To give a better presentation of our findings, the remaining of this paper is arranged as follows. Section \Rmnum{2} formulates the merging problem at highway on-ramps. Section \Rmnum{3} presents three cooperative driving strategies. Section \Rmnum{4} gives a brief analysis of the grouping based strategy. Section \Rmnum{5} provides the simulation results of several experiments to validate the effectiveness of the proposed strategy. Finally, concluding remarks are given in Section \Rmnum{6}.

\section{Problem Formulation}

\begin{figure}[htb]
    \centering
    \includegraphics[width=7cm]{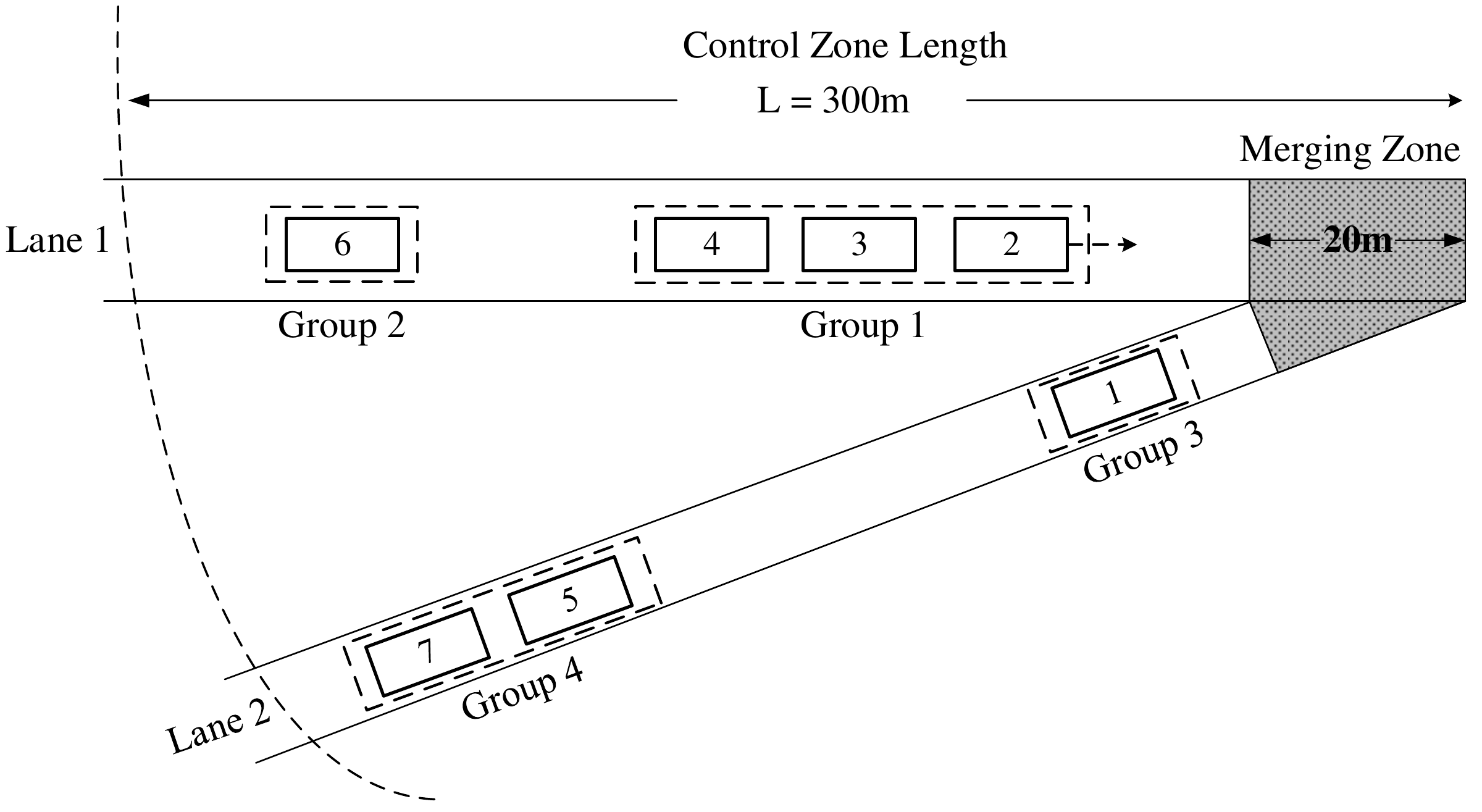}
    \caption{A typical merging scenario at a highway on-ramp.}
    \label{fig:scenario}
\end{figure}

This paper considers a highway on-ramp with a single lane in each movement as shown in Fig.\ref{fig:scenario}. The shadow area is called as merging zone where two CAVs on different movements may collide. $L$ is the distance from the entry of control zone to the merging zone.
Usually, $L$ is about $50$m-$200$m.
The multiple lane merging scenario is similar and the proposed strategy also can be employed with slight modification\cite{li2005cooperative,li2006cooperative}.
Tab.\ref{tab:Mainnotation} gives the nomenclature list of the major symbols used in this paper.

Cooperative driving strategy aims to schedule the velocity and acceleration profiles of all CAVs \cite{zhang2016optimal,malikopoulos2017decentralized}.
As pointed out in \cite{meng2017analysis}, the performance of a strategy mainly depends on the passing order of vehicles and the differences between different motion planning methods are negligible.
Thus, the merging problem is transferred into an optimization problem with respect to the passing order together with a simple motion planning method which requires little computational cost.
In this paper, we will focus on the first problem.
The details of the motion planning used in this paper are presented in \textbf{Appendix A}.

Once a CAV enters into the control zone, it is given a unique identity.
CAV$_i$ means it is the $i$th CAV that enters into the control zone.
The movements of each vehicle within the control zone may be re-scheduled for every $T$ minutes.
Every time the schedule begins, the objective of the optimization problem can be written as

\begin{equation}
J=\omega_1max(t_{assign,i})+\omega_2\sum_{i=1}^n(t_{assign,i}-t_{min,i}),
\end{equation}

\noindent where $t_{assign,i}$ is the decision variable and represents the desired access time to the merging zone for CAV$_i$. $n$ is the total number of vehicles in the control zone. $t_{min,i}$ is the minimum access time to the merging zone and can be easily derived by

\begin{subequations}
\begin{align}
&t_{min}=t_0+t_1+t_2,\\
&v=\sqrt{v_0^2+2a_{max}x_0},\\
&t_1=min\bigg(\frac{v_{max}-v_0}{a_{max}},\frac{v-v_0}{a_{max}}\bigg),\\
&t_2=max\bigg(\frac{2a_{max}x_0-v_{max}^2+v_0^2}{2a_{max}v_{max}},0\bigg),
\end{align}
\label{eq:tmin}
\end{subequations}

\noindent where $x_0$ is initial location, $v_0$ is initial velocity, and $t_0$ is the time when the CAV enters into to the control zone.

The first term in the objective is to minimize passing time and the second term is to decrease the delay of CAVs. $\omega_1$ and $\omega_2$ are weighted parameters of the objective.

Suppose that CAV$_i$ and CAV$_{i+1}$ are two consecutive CAVs on the same movement. To avoid a rear-end collision, we require that the minimum allowable safety gap between them is larger than $\Delta t_1$

\begin{equation}
t_{assign,i}-t_{assign,i+1}\geq \Delta t_1.
\end{equation}

Suppose that CAV$_i$ and CAV$_{j}$ are two CAVs on conflict movements. To avoid a lateral collision, we require the minimum allowable safety gap between them is larger than $\Delta t_2$

\begin{equation}
\begin{split}
&t_{assign,i}-t_{assign,j}\geq \Delta t_2,\\
&OR\\
&t_{assign,j}-t_{assign,i}\geq \Delta t_2.\\
\end{split}
\end{equation}

\noindent The constraints ensure that for any two CAVs $i$ and $j$ that are on the conflict movements, only one CAV can enter into the merging zone after the other CAV has left the merging zone. Moreover, we assume that $\Delta t_2$ is greater than $\Delta t_1$.

Introducing some binary variables, we can formulate the whole optimization problem in terms of the passing order (decision variable) $\bm{b}=[b_{k_1,l_1},\cdots,b_{k,l},\cdots,b_{k_{n_1},l_{n_2}}] \in \left\{0, 1\right\}^{n_1 \times n_2}$ as below

\begin{subequations}
\begin{align}
\mathop{\min}_{\bm{t_{assign}},\bm{b}}\quad &\omega_1max(t_{assign,i})+\omega_2\sum_{i=1}^n(t_{assign,i}-t_{min,i})\\
\text{subject to } & a_{min} \leq a_i \leq a_{max} \label{eq:optimization_d}\\
& v_{min} \leq v_i \leq v_{max} \label{eq:optimization_e}\\
&t_{assign,i}-t_{assign,j}\geq \Delta t_1 \label{eq:optimization_a}\\
&t_{assign,k}-t_{assign,l}+M\cdot b_{k,l}\geq \Delta t_2 \label{eq:optimization_b}\\
&t_{assign,l}-t_{assign,k}+M\cdot (1-b_{k,l})\geq \Delta t_2 \label{eq:optimization_c}\\
&k\in \mathbb{N}_1=\{k_1,k_2,\cdots,k_{n_1}\}\\
&l\in \mathbb{N}_2=\{l_1,l_2,\cdots,l_{n_2}\}\\
&b_{k,l} \in \{0,1\}
\end{align}
\label{eq:optimization}
\end{subequations}

\noindent where $M$ is a sufficiently big number, $\mathbb{N}_1$ and $\mathbb{N}_2$ are two sets that contain all CAVs travelling on two movements. The sizes of $\mathbb{N}_1$ and $\mathbb{N}_2$ are $n_1$ and $n_2$ respectively. $b_{k,l}$ is a binary number. When $b_{k,l}$ equals 0, CAV$_k$ passes through the merging zone earlier than CAV$_l$.

Similar to \cite{li2005cooperative}, the passing order also can be denoted as a string, which is more intuitive in the analysis.
For example, string \textbf{ABCD} means CAV$_A$, CAV$_B$, CAV$_C$ and CAV$_D$ enter the merging zone sequentially.
Each such a string corresponds to a possible value of $\bm{b}$.

If the passing order is given, the problem (\ref{eq:optimization}) can be solved by a simple iteration algorithm shown in \textbf{Algorithm 1}.
\begin{algorithm}
\small
\caption{A Simple Iteration Algorithm}
\label{assigned_time}
\begin{algorithmic}[1]
\begin{spacing}{1.2}
\REQUIRE A passing order $P$
\ENSURE An objective value $J$ and $t_{assign}$
\STATE $t_{assign}(1)=\text{CAV}_{P,1}[t_{min}]$
\FOR {each $i \in [2,length(P)]$}
\IF{CAV$_{P,i-1}$ and CAV$_{P,i}$ are on the same movement}
\STATE
$t_{assign}(i)=max(t_{assign}(i-1)+\Delta t_1,\text{CAV}_{P,i}[t_{min}])$
\ELSE
\STATE
$t_{assign}(i)=max(t_{assign}(i-1)+\Delta t_2,\text{CAV}_{P,i}[t_{min}])$
\ENDIF
\ENDFOR
\STATE $J=\omega_1max(t_{assign})+\omega_2\sum_{i=1}^{length(P)}(t_{assign}(i)-\text{CAV}_{P,i}[t_{min}])$
\end{spacing}
\end{algorithmic}
\end{algorithm}

Here, $\text{CAV}[t_{min}]$ means the minimum access time $t_{min}$ of the CAV. CAV$_{P,i}$ is the $i$th CAV passing through the merging zone in the passing order $P$.

Obviously, the time complexity of the \textbf{Algorithm 1} is $O(n_1+ n_2)$.

\section{Cooperative Driving Strategies}

In this section, we will present three cooperative driving strategies to solve the above optimization problem (\ref{eq:optimization}).

\subsection{Planning Based Strategy}

Generally, planning based strategy directly attacks the above mixed integer linear programming (MILP) problem (\ref{eq:optimization}). We can use either tree-based enumeration method \cite{li2006cooperative,meng2017analysis} or classic branch-and-bound method to solve this MILP \cite{linderoth2010milp,he2014learning}

However, there are $2^{n_1n_2}$ possible values for variable $\bm{b}$. So, the time complexity of branch-and-bound method is still exponential in the worst case.
Numerical tests show that the enumeration based method only works well when the number of vehicles is less than 12 \cite{meng2017analysis}.
The time efficiency of the branch-and-bound method is similar.

\subsection{Ad Hoc Negotiation Based Strategy}

Ad hoc negotiation based strategy uses greedy search to solve Problem (\ref{eq:optimization}). As summarized in \cite{dresner2004multiagent,dresner2008multiagent}, the passing order in many ad hoc negotiation methods follows the first-in-first-out (FIFO) principle. In other words, all CAVs in the control zone estimate their arrival time points to the merging zone if no schedule is given. The passing order is derived by sorting their arrival time points in ascending order.

When the passing order is determined, the degenerated problem is easily solved by \textbf{Algorithm 1}.
It is easy to show that the time complexity of ad hoc negotiation based strategy is $O(n_1 +n_2)$.

\subsection{Grouping Based Strategy}

Grouping based strategy can be viewed as a modification of planning based strategy.
Its main idea is to search the sub-optimal passing order among a subset of search space instead of all passing orders.
This subset is determined by the following grouping method: if the headway between two consecutive vehicles is smaller than a grouping threshold, then they will be grouped into the same group.
The vehicles in a group are assumed to enter the merging zone consecutively without any other vehicles' interruption.

The grouping threshold determines the number of groups.
To control the time consumption, the maximum allowable number of groups is set as 12 in this paper.
If the number of groups is larger than 12, the time consumption will be too large for practical applications.
Fortunately, we do not need to consider too many vehicles for merging problems.
So, 12 groups usually meet our expectation to balance the complexity and efficiency of the planning algorithm.

Moreover, we apply an adaptive threshold in this paper.
The initial value of the threshold is set as 1.5m which is the minimum safety gap between two consecutive vehicles.
If all vehicles had been grouped into less than 12 groups, we stop.
Otherwise, we increase the threshold for 0.1s each time and re-group the vehicle in an iteration manner, until the number of groups is not larger than 12.

When grouping is done, we consider each group as a special CAV and calculate the optimal passing order for these special CAVs.
Finally, the obtained passing order for the groups of vehicles will be interpreted into the passing order for all vehicles.

The major benefit of grouping is to reduce the time complexity of the problem.
If the maximum number of groups is $c$, the time complexity of the grouping based strategy is approximately $O(c!\cdot (n_1 + n_2))$.
This greatly saves the planning time, especially when $(n_1 + n_2)$ is large.

\begin{figure}[htb]
    \centering
    \includegraphics[width=9cm]{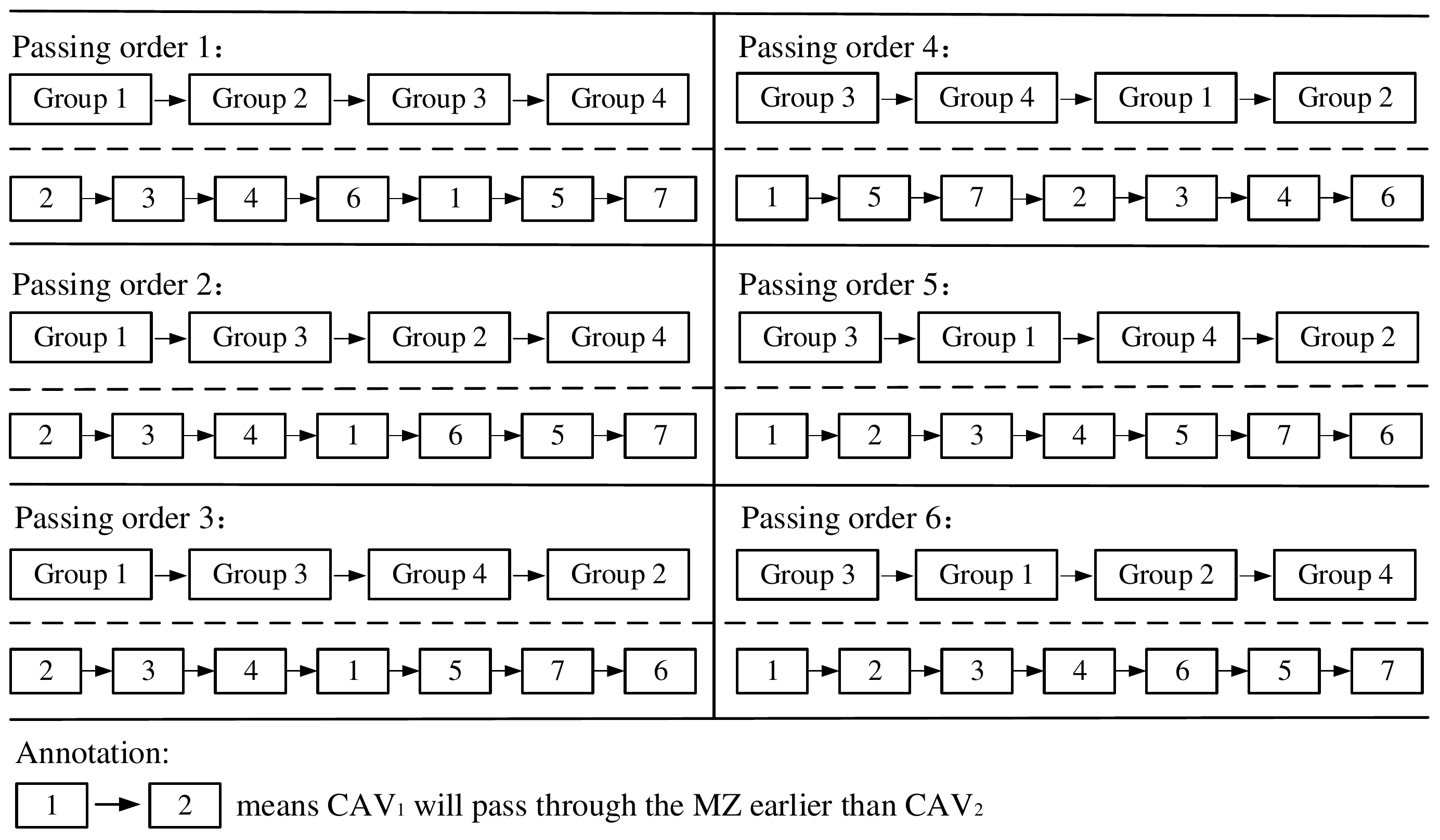}
    \caption{The possible passing orders for the grouping based strategy. An order of groups corresponds to a passing order of CAVs.}
    \label{fig:subset}
\end{figure}

To better understand the benefit of the proposed strategy, we briefly introduce some typical cases for examples. As shown in Fig.\ref{fig:scenario} as an example. 7 vehicles are grouped into 4 groups. As shown in Fig.\ref{fig:subset}, the number of searched passing orders is largely reduced from $7!$ to 6.

In the rest of this paper, we will show that the sub-optimal passing order found by the grouping based strategy has a high probability to be close to the global optimal passing order, if the grouping threshold is appropriately chosen.

\section{A Brief Analysis}

In this section, we will show that it is usually unnecessary to divide two consecutive vehicles whose intervehicle headway is very small apart and let other vehicles cut in.
More precisely, we study a very basic scenario in which we can enumerate all the candidate passing orders.
We compare the traffic efficiency in each possible situation with and without grouping.
We will show that the occurring probability for the special case (in which grouping leads to a larger passing time than not grouping) is very small.

\begin{figure}[htb]
    \centering
    \includegraphics[width=7cm]{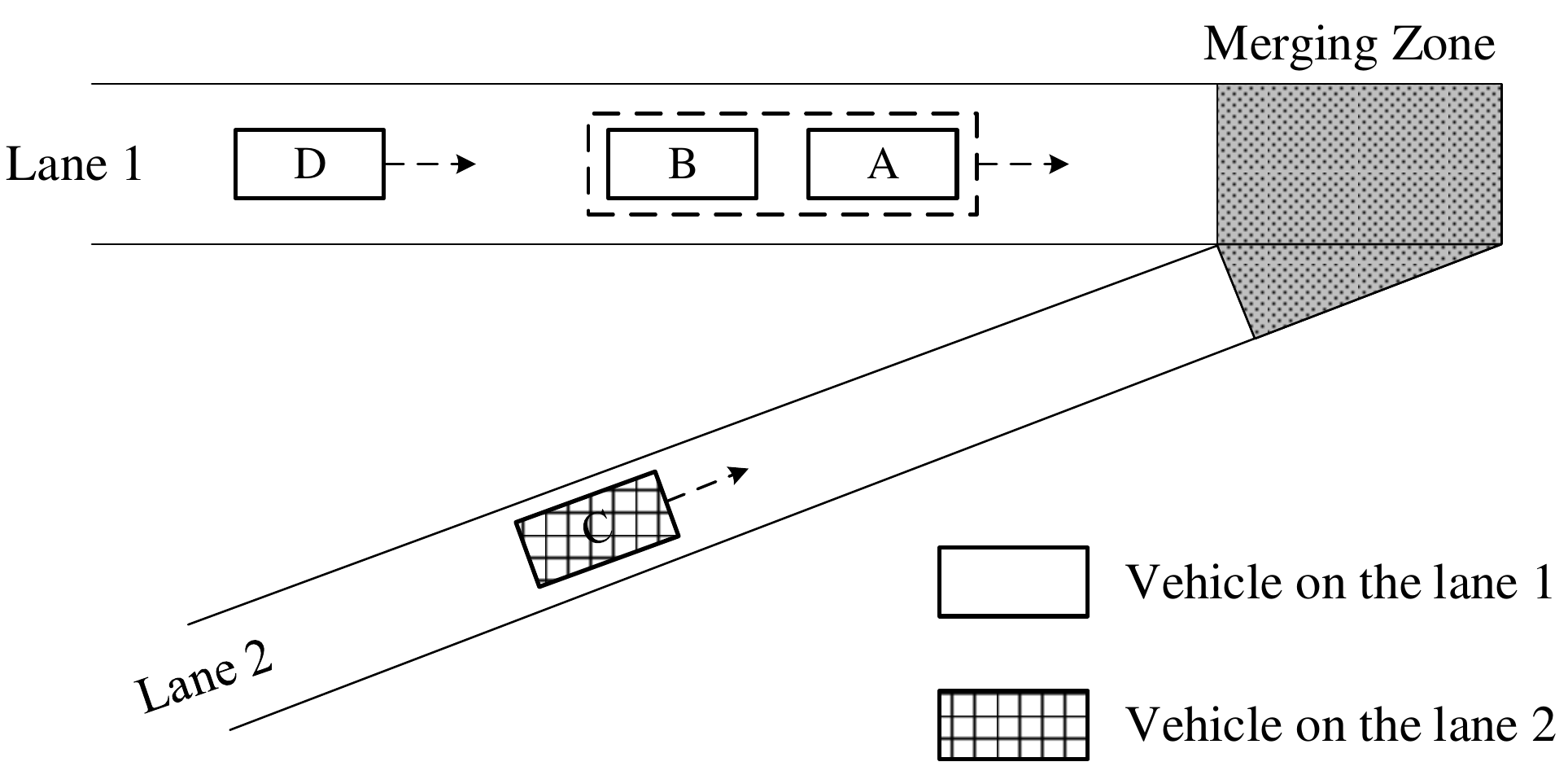}
    \caption{A merging scenario.}
    \label{fig:grouping}
\end{figure}

To this end, let us consider the following scenario that consists of four CAVs. CAV$_A$, CAV$_B$, and CAV$_D$ are on lane 1; and CAV$_C$ is on lane 2; see Fig.\ref{fig:grouping}.
The headway between CAV$_A$ and CAV$_B$ is less than the grouping threshold, so they may be in a group. For convenience, we denote the initial headway between CAV$_i$ and CAV$_j$ as $h_{i,j}$, when these vehicles enter the control zone.
Also, we have the reciprocal relationship between the average headway and the arrival rate $\lambda$ of vehicles.

\begin{figure}[htb]
    \centering
    \includegraphics[width=8cm]{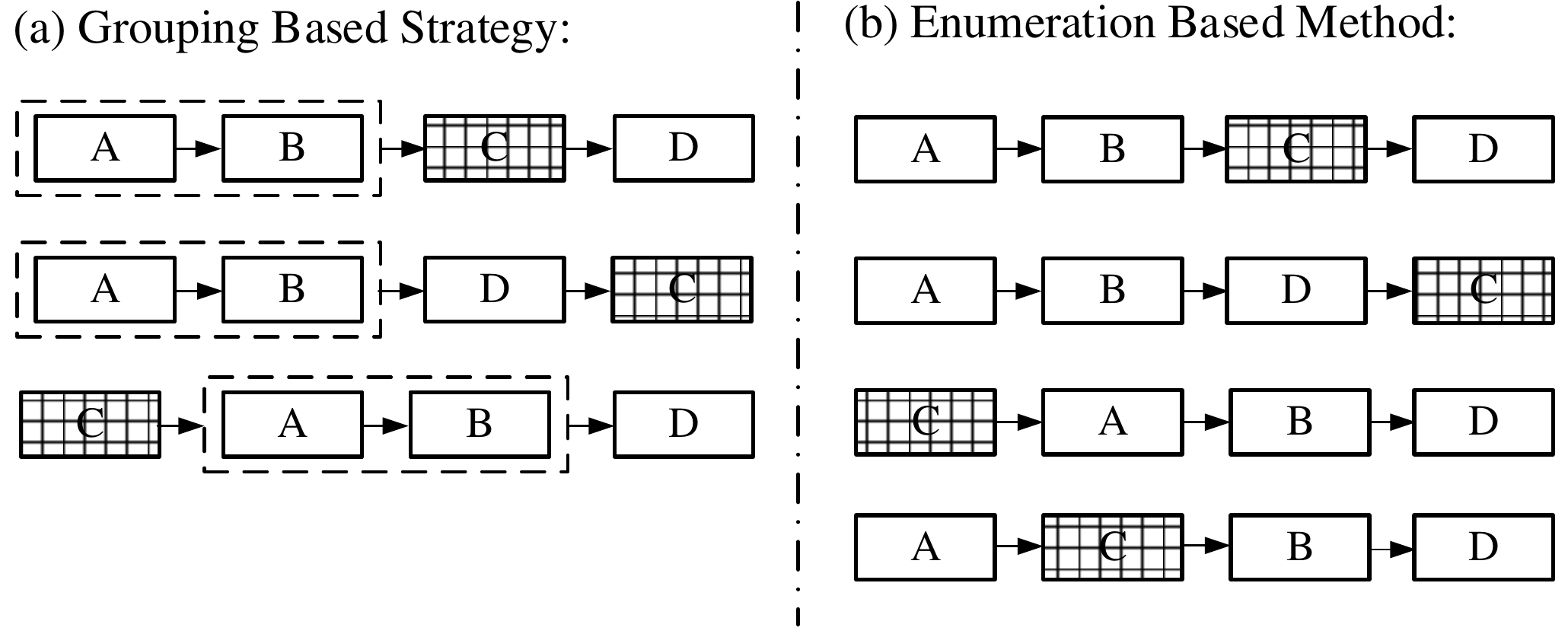}
    \caption{(a) All 3 passing orders after grouping. (b) All 4 passing orders without grouping.}
    \label{fig:solution_comparison}
\end{figure}

All valid passing orders with and without grouping are shown in Fig.\ref{fig:solution_comparison}.
As aforementioned, we use a string to denote the passing order of vehicle for presentation simplicity.

Obviously, for this scenario, if the global optimal solution (passing order) is \textbf{ABCD}, \textbf{ABDC}, or \textbf{CABD}, grouping does not hinder us to find the global optimal solution.
The only special case that we should take care is the global optimal solution is \textbf{ACBD}.

In the rest, we will discuss when \textbf{ACBD} can be the global optimal solution.

The optimal solution is \textbf{ACBD} means that we have $t_{ACBD}\le t_{ABCD}$, $t_{ACBD}\le t_{CABD}$ and $t_{ACBD}\le t_{ABDC}$ simultaneously.
Let us take $t_{ACBD}\le t_{ABCD}$ as an example to show what constraints are needed.

If the passing order is \textbf{ACBD}, the assigned time of each vehicle can be derived according to \textbf{Algorithm 1}.

\begin{subequations}
\begin{align}
t_{assign,A}&=t_{min,A},\\
t_{assign,C}&=max\{t_{assign,A}+\Delta t_2, t_{min,C}\},\\
t_{assign,B}&=max\{t_{assign,C}+\Delta t_2, t_{min,B}\},\\
t_{assign,D}&=max\{t_{assign,B}+\Delta t_1, t_{min,D}\}.
\end{align}
\end{subequations}

Then, the total passing time of all 4 vehicles under the passing order \textbf{ACBD} is

\begin{equation}
\begin{split}
t_{ACBD}=&t_{assign,D}\\
=&max\{t_{min,A}+\Delta t_1+2\Delta t_2, t_{min,B}+\Delta t_1, \\
&t_{min,C}+\Delta t_1+\Delta t_2, t_{min,D}\}.
\label{eq:passingorderACBD}
\end{split}
\end{equation}

Similarly, the total passing time of all 4 vehicles under the passing order \textbf{ABCD} is

\begin{equation}
\begin{split}
t_{ABCD}=&max\{t_{min,A}+\Delta t_1+2\Delta t_2, t_{min,B}+2\Delta t_2, \\
& t_{min,C}+\Delta t_2, t_{min,D}\}.
\label{eq:passingorderABCD}
\end{split}
\end{equation}

For each case, we will compare the value of $t_{ACBD}$ and $t_{ABCD}$. The following analysis shows that only in the fourth case, $t_{ACBD}$ can be smaller than $t_{ABCD}$.

\begin{enumerate}

\item if $t_{ABCD}=t_{min,A}+\Delta t_1+2\Delta t_2$,
    \begin{equation}
        t_{ACBD}-t_{ABCD} \geq t_{min,A}+\Delta t_1+2\Delta t_2 - t_{ABCD} = 0.
    \label{eq:ABC1}
    \end{equation}

\item if $t_{ABCD}=t_{min,C}+\Delta t_2$,
    \begin{equation}
        t_{ACBD}-t_{ABCD} \geq t_{min,C}+\Delta t_1+\Delta t_2 - t_{ABCD} = \Delta t_1.
    \label{eq:ABC2}
    \end{equation}

\item if $t_{ABCD}=t_{min,D}$,
    \begin{equation}
        t_{ACBD}-t_{ABCD} \geq t_{min,D} - t_{ABCD} = 0.
    \label{eq:ABC3}
    \end{equation}

\item if $t_{ABCD}=t_{min,B}+2\Delta t_2$ and $t_{ABCD} > t_{min,C}+\Delta t_1+\Delta t_2$,
    \begin{equation}
        t_{ACBD}-t_{ABCD} < 0.
    \label{eq:ABC4}
    \end{equation}

\end{enumerate}

Therefore, the special case (e.g. $t_{ACBD}-t_{ABCD} < 0$) only occurs when the following constraints are satisfied.

\begin{equation}
\begin{cases}
t_{min,C}\le t_{min,B}+\Delta t_2 - \Delta t_1\\
t_{min,A}+\Delta t_1 \leq t_{min,B} \\
t_{min,D} \le t_{min,B}+2\Delta t_2
\end{cases}
\label{eq:constraints1}
\end{equation}

The constraints for $t_{ACBD}\le t_{CABD}$ and $t_{ACBD}\le t_{ABDC}$ can be derived by the similar way.
Summarizing all the constraints and relaxing some of them, we get
\begin{equation}
    \begin{cases}
    \Delta t_1 \leq h_{A,B} \leq \delta\\
    -\Delta t_1 + \Delta t_2 \le t_{min,C}-t_{min,A} \le -\Delta t_1 + \Delta t_2 +\delta\\
    \Delta t_1+ \Delta t_2 \le t_{min,D}-t_{min,A} \le 2\Delta t_2 + \delta
    \end{cases}
    \label{eq:constraints3}
\end{equation}

Finally, we discuss the relationship between $t_{min,j}-t_{min,i}$ and the headway $h_{i,j}$ to check the occurring probability of such a case.
For simplicity, we assume that all CAVs are operated at the maximum velocity and thus have $t_{min,j}-t_{min,i}=h_{i,j}$.

As suggested in \cite{korkmaz2010supporting}, we suppose that the headway $h$ follows displaced exponential distribution as

\begin{equation}
f(h)=\frac{1}{\bar{h}-\tau}e^{-(h-\tau)/(\bar{h}-\tau)},\quad h\geq\tau,
\end{equation}

\noindent where $\bar{h}=1/\lambda$ is the average headway, $\lambda$ is the average arrival rate, $\tau$ is the minimum headway which equals $\Delta t_1$ in the paper.
The cumulative distribution function of the headway is

\begin{equation}
P(h\le H)=F(H)=1-e^{-(H-\Delta t_1)/(\bar{h}-\Delta t_1)}.
\label{eq:cdf}
\end{equation}
\noindent For presentation convenience, we define $F(H_1,H_2)=P(H_1\le h\le H_2)=F(H_2)-F(H_1)$.

According to (\ref{eq:cdf}), the probability of satisfying the constraints (\ref{eq:constraints3}) is

\begin{equation}
P=F_1(\Delta t_1,\delta)\cdot F_2(\Delta t_2-\Delta t_1,\Delta t_2-\Delta t_1 +\delta)\cdot F_1(\Delta t_1+ \Delta t_2,2\Delta t_2 + \delta).
\end{equation}

When we set $\Delta t_1 =1.5$s, $\delta=2$s, $\Delta t_2=2.5$s, and $\lambda_1=\lambda_2=0.2$veh/s, this probability is about 1.3\%. That is, under this parameter setting, the probability that the sub-optimal solution of the grouping based strategy is worse than the global optimal solution is 1.3\%.
This is really a small chance.

The above analysis reveals the major reason why the sub-optimal solution of the grouping based strategy is good enough in most situations.
However, this method becomes improbable to analyze the cases that consist of a lot of vehicles, since the number of possible cases increases exponentially.
So, in the rest of this paper, we resort to numerical tests to further validate our conclusion.

\section{Simulation Tests}

The first experiment introduce the concept of alignment probability initialized in ordinal optimization \cite{ho2008ordinal,ho1999explanation,lau1997universal} as the measure to validate that the sub-optimal solution found by grouping method is a good enough solution.
We also use the histogram of all possible solutions (passing orders) of the merging problem to show that the grouping based strategy obtains a good enough solution.
The second experiment shows that the time consumption of the grouping based strategy is much less than conventional planning based methods.
Finally, the third experiment compares the performance of different cooperative driving strategies.

In these experiments, the vehicles' arrival at each movement is assumed to be a Poisson process. The arrival rate can be varied to test the performance of the proposed strategy under different traffic demands. The vehicles' arrival rates at two movements are the same unless otherwise specified.

The weight parameters of objective function $\omega_1$ and $\omega_2$ both are 0.5. The minimum safety gaps of three strategies are the same. The minimum safety gap between two CAVs on the same movement is 1.5s, and the minimum safety gap between two CAVs on the conflict movements is 2s.

All experiments are carried out on a MATLAB platform in a personal computer with an Intel i7 CPU and an 8GB RAM.

\subsection{The Optimality of the Grouping Based Strategy}

The grouping based strategy can be regarded as a sampling technique to narrow down the search space and speeding the search process.
Alignment probability is a nice measure to characterize the degree of matching between the original solution space $S$ and the sampled subset $G$ \cite{lau1997universal}.

Suppose $G$ is a good enough set which consists of the top-$g$ solutions of search space $\bm{B}$. $g$ is the ranking index. For example, the top-$1$ solution denotes the best solution. $S$ is a selected set where the members are selected by using certain sampling technique or selection rule. $|G\cap S|\geq k^\prime$ means there are at least $k^\prime$ truly good solutions in $S$. $k^\prime$ is called the alignment level and $P(|G\cap S|\geq k^\prime)$ is called alignment probability\cite{ho2008ordinal,ho1999explanation}.

In this paper, the alignment probability is calculated through simulation experiments.
In the experiment, we vary $\lambda$ from 0.1 veh/(lane$\cdot$s) to 0.25 veh/(lane$\cdot$s). Under each certain arrival rate, we simulate a 20 minutes traffic process. We record all solutions of the enumeration based method, the estimated optimal solution of the grouping based strategy, time consumption, and the number of vehicles in the merging zone.

\begin{figure}[htbp]
    \centering
    \includegraphics[width=7cm]{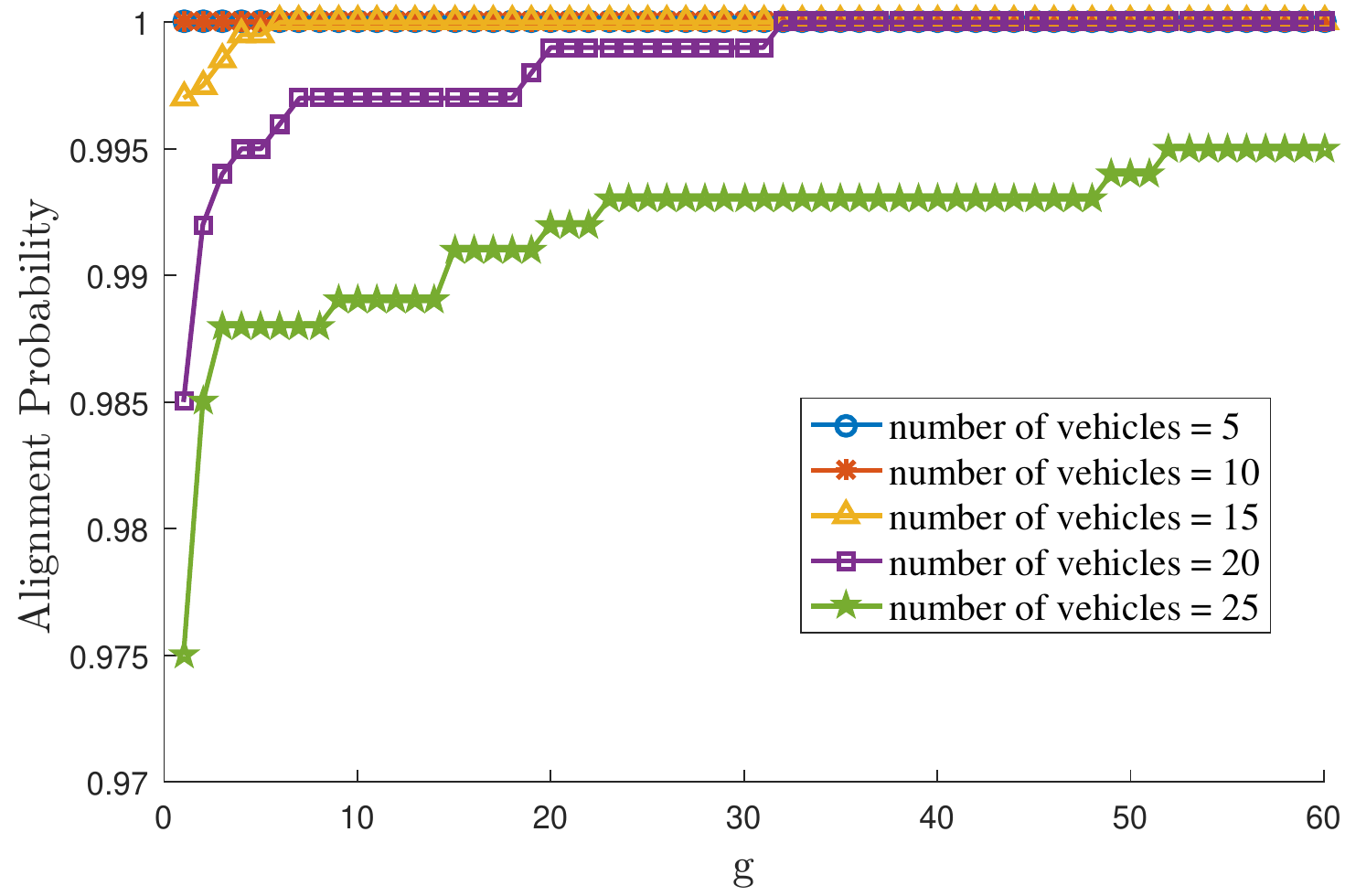}
    \caption{Alignment probability versus top-$g$ solutions parameterized by the number of vehicles.}
    \label{fig:alignment_probability}
\end{figure}

We compare the estimated optimal solution with top-$g$ solutions to count the alignment probability.
We set the alignment level $k^\prime$ as 1, since we care about whether there is a good enough solution in the selected set.
The alignment probabilities with respect to different numbers of vehicles are shown in Fig.\ref{fig:alignment_probability}.

It is clear that the sub-optimal solution always is among the top-$0.025\%$ solutions and is good enough.
Even when the number of vehicles equals 20 and the average number of the possible solutions is about 127000, the sub-optimal solution found by grouping method is among the top-32 solutions.
In other words, from the viewpoint of solutions' order, the sub-optimal solution can be very close to the global optimal solution with a high probability.

\begin{figure}[htbp]
    \centering
    \includegraphics[width=8.5cm]{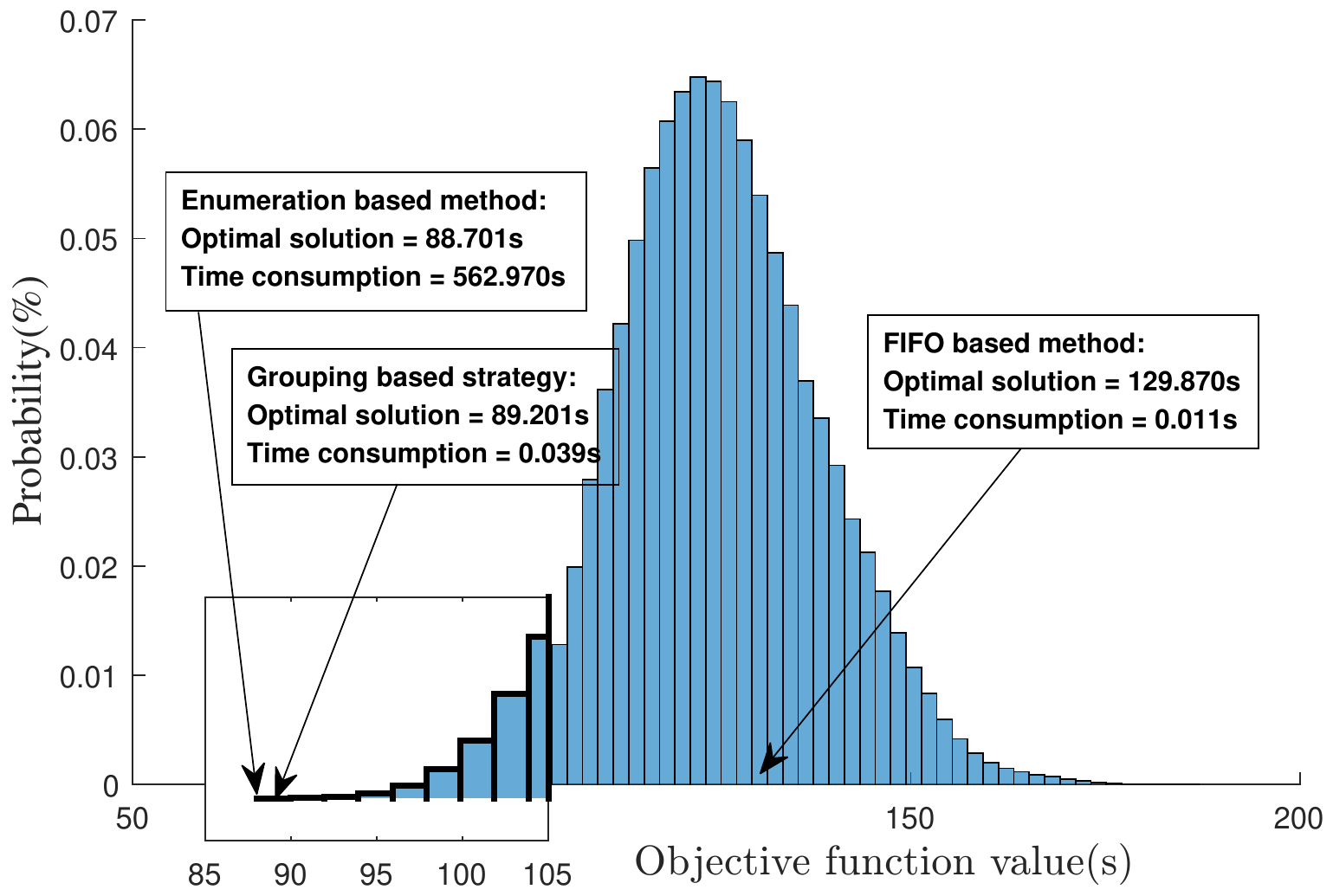}
    \caption{The histogram of solution values for a merging scenario with 25 CAVs.}
    \label{fig:solution}
\end{figure}

For a special merging scenario, we can calculate all the objective values for all the possible solutions (passing orders) and plot them in a histogram manner. This histogram intuitively describes the performance of solutions.

Fig.\ref{fig:solution} gives such a histogram for solution values for a merging scenario with 25 CAVs.
There are 5200300 possible passing orders for the merging scenario.

We apply the FIFO based ad hoc negotiation strategy and the grouping based strategy for the same scenario. Then, we mark the locations of their optimal solutions in the Fig.\ref{fig:solution}.
It is clear that the solution found by the grouping based strategy is nearly the same as the global optimal solution; while the solution found by the FIFO based ad hoc negotiation strategy is far away from the global optimal solution. Indeed, the solution found by the grouping based strategy is the 7th best solution; while the solution of the FIFO based strategy ranks 3350239th.

\begin{table*}[htbp]
  \centering
  \small
  \caption{Comparison results of three cooperative strategies}
    \begin{tabular}{ccccc}
    \toprule
    \multirow{2}[2]{*}{average arrival rate$^1$} & \multirow{2}[2]{*}{strategies} & \multirow{2}[2]{*}{average delay(s)} & \multirow{2}[2]{*}{average time consumption(ms)} & \multicolumn{1}{c}{\multirow{2}[2]{*}{average number of CAVs}} \\
          &       &       &       &  \\
    \midrule
    \multirow{3}[2]{*}{0.1} & Grouping based & 0.574 & 1.70  & 6.44 \\
          & Planning based & 0.574 & 101.70 & 6.44 \\
          & Ad hoc negotiation based & 0.602 & 0.08  & 6.45 \\
    \midrule
    \multirow{3}[2]{*}{0.15} & Grouping based & 0.759 & 12.20 & 9.51 \\
          & Planning based & 0.759 & 230.00 & 9.51 \\
          & Ad hoc negotiation based & 0.913 & 0.10  & 9.58 \\
    \midrule
    \multirow{3}[2]{*}{0.2} & Grouping based & 1.046 & 18.50 & 12.11 \\
          & Planning based & 1.046 & 1310.70 & 12.11 \\
          & Ad hoc negotiation based & 1.568 & 0.12  & 12.31 \\
    \midrule
    \multirow{3}[2]{*}{0.25} & Grouping based & 1.984 & 34.90 & 15.97 \\
          & Planning based & 1.942 & 9055.20 &  15.86 \\
          & Ad hoc negotiation based & 4.235 & 0.10  & 16.55 \\
    \midrule
    \multirow{3}[2]{*}{0.3$^2$} & Grouping based & 2.741 & 37.70 & 18.76 \\
          & Planning based & 2.726 & 11464.80 & 18.73 \\
          & Ad hoc negotiation based & 5.722 & 0.10  & 19.94 \\
    \bottomrule
    \end{tabular}%
    \begin{tablenotes}
            \footnotesize
            \item{$^1$} Suppose average arrival rates on two movements are the same.
            \item{$^2$} When the arrival rates on two movements are 0.3, the traffic becomes seriously congested and most CAVs block in the upstream after 5-minute simulation. Thus, the results in this parameter setting are the results of a 5-minute simulation.
    \end{tablenotes}
  \label{tab:result3}%
\end{table*}%

\subsection{The Time Consumption of the Grouping Based Strategy}

To check the average time consumption with respect to different numbers of vehicles that will be considered, we vary the vehicle arrival rate $\lambda$ from 0.1 veh/(lane$\cdot$s) to 0.25 veh/(lane$\cdot$s).
Under each certain arrival rate, we simulate a 20 minutes traffic process.
We record corresponding time consumption and the number of vehicles in the merging zone.

\begin{figure}[htbp]
    \centering
    \includegraphics[width=7.5cm]{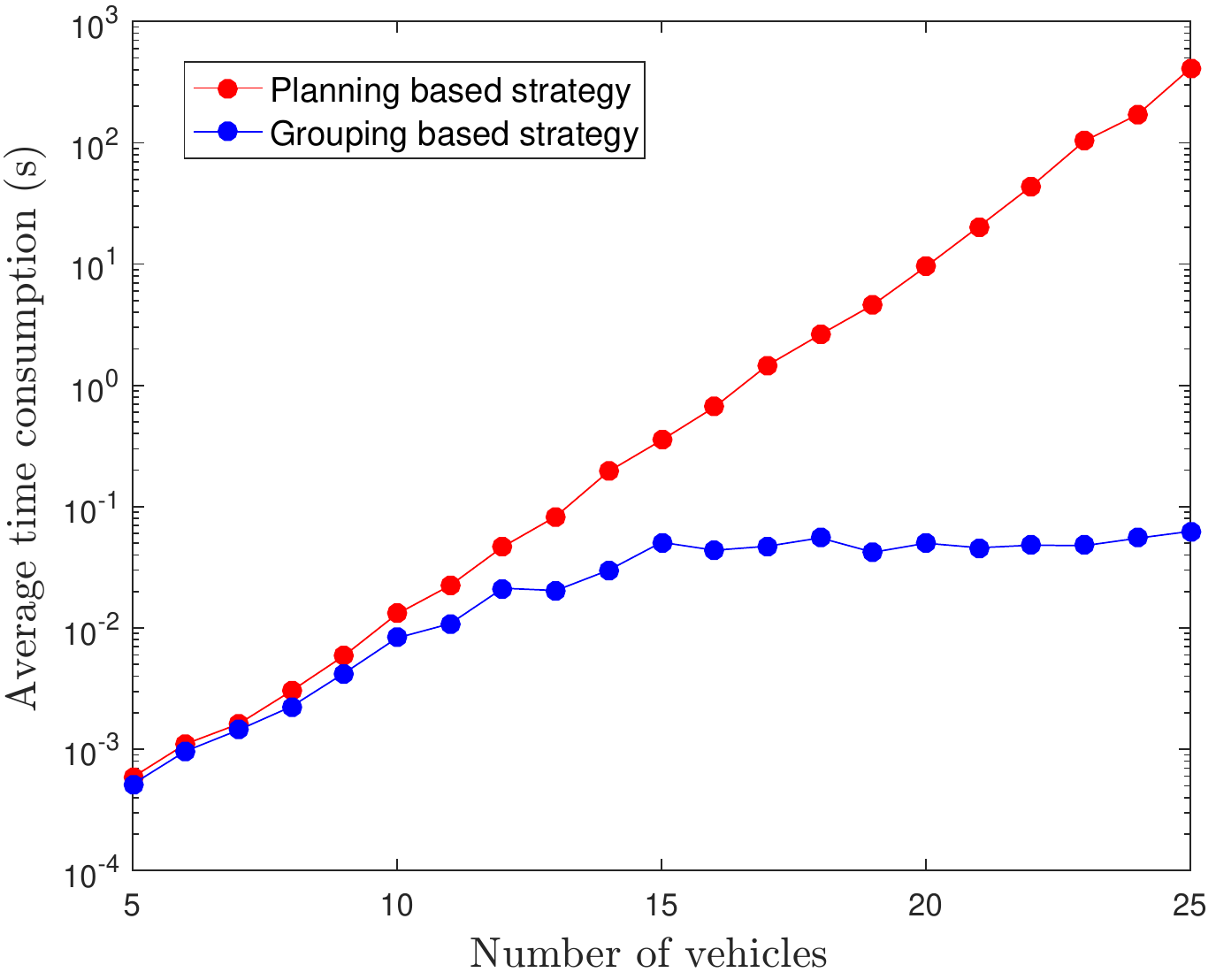}
    \caption{A semi-log plot of average time consumption against the number of vehicles. The result comes from 4800 merging scenarios.}
    \label{fig:time_consumption}
\end{figure}

As shown in Fig.\ref{fig:time_consumption}, the results indicate that the average time consumption of the planning based strategy (enumeration based method) increases almost exponentially.
Although the planning based strategy gives the global optimal solution, it is difficult to be applied in practice.
In contrast, since we restrict the maximum group number is 12, the average time consumption of the grouping based strategy reaches a saturated value, when the number of vehicles is larger than 15.

Combining Fig.\ref{fig:alignment_probability} and Fig.\ref{fig:time_consumption}, we can conclude that the grouping method can find a good enough solution within a short enough time.

\subsection{A Comparison of Different Cooperative Driving Strategies with Respect to Arrival Flow Rate}

To compare the performance of different cooperative driving strategies, we calculate the delay of CAV$_i$ as

\begin{equation}
t_{delay,i}=t_{assign,i}-t_{min,i}.
\end{equation}

\noindent and the time consumption that a strategy takes to find the passing order.
A less delay indicates that a better performance; and a less time consumption indicates that a faster speed.

In the simulation, the initial merging scenario contains five CAVs whose initial positions are generated randomly. The average arrival rate $\lambda$ of the following CAVs can vary from 0.1 veh/(lane$\cdot$s) to 0.3 veh/(lane$\cdot$s). For each arrival rate, we simulate a 20 minutes traffic process.

Grouping based strategy, planning based strategy (branch and bound method), and ad hoc negotiation based strategy (FIFO based method) are respectively applied to the same initial merging scenario.
All the considered CAVs' trajectories are replanned every $T =2$ seconds.

In planning based strategy, the MILP problem (\ref{eq:optimization}) is directly solved by CVX software with Mosek solver. The Mosek solver makes use of branch and bound method to handle integer variables \cite{linderoth2010milp}. The performance measures are shown in Tab.\ref{tab:result3}.

As shown in Tab.\ref{tab:result3}, the average delay of the grouping based strategy and planning based strategy is similar and the biggest difference is only 0.04 s/veh. However, the time consumption of the planning based strategy increases sharply with the arrival rate. When the arrival rate equals 0.3, the average time consumption of the planning based strategy is more than 11 seconds. At the same time, the calculation of the proposed strategy always can be finished within 40 ms.

On the other hand, under the situation of high arrival rate, the coordination performance of the ad hoc negotiation based strategy is extremely bad. When the arrival rate equals 0.25, the average delay of the ad hoc negotiation based strategy is more than twice that of other two strategies. Although the ad hoc negotiation based strategy is time-saving, its performance is far from satisfactory.

It is obvious that the grouping based strategy makes a good tradeoff between the planning based strategy and ad hoc negotiation based strategy. Its good traffic control performance and short time consumption make it practical for real applications.

\section{Conclusion}

In this paper, we propose a grouping based strategy to make a good tradeoff between computation complexity and traffic control efficiency.
Its key idea is to narrow down the number of candidate passing orders and search a sub-optimal passing order among a subset of the solution space. Analysis and simulation results validate that the sub-optimal solution found by the grouping based strategy has a high probability to be close to the global optimal solution, no matter what vehicle arrival rate is given.
Being compared with the planning based strategy, the grouping based strategy yields similar traffic efficiency with much less calculation time.
Being compared with ad hoc negotiation based strategy, the grouping based strategy gives much better traffic efficiency with similar calculation time.
Thus, we recommend the grouping based strategy as a promising cooperative driving strategy in practice.

It should be pointed out that the empirical dynamic constraints of vehicles are not considered in this paper.
We are currently building several automated vehicle prototypes.
In the near future, we will test our grouping based strategy in a real on-ramp and design new tracking controllers to implement the planned trajectories.

\appendices
\section{A Simple Motion Planning Method}

This paper uses a similar motion planning method as used in \cite{fayazi2017optimal}. The method can be easily derived by basic kinematics and requires little computational cost.

\newsavebox\unionef
\begin{lrbox}{\unionef}
    \setlength{\fboxrule}{0pt}
  \begin{minipage}{4cm}
    \includegraphics[width=4cm]{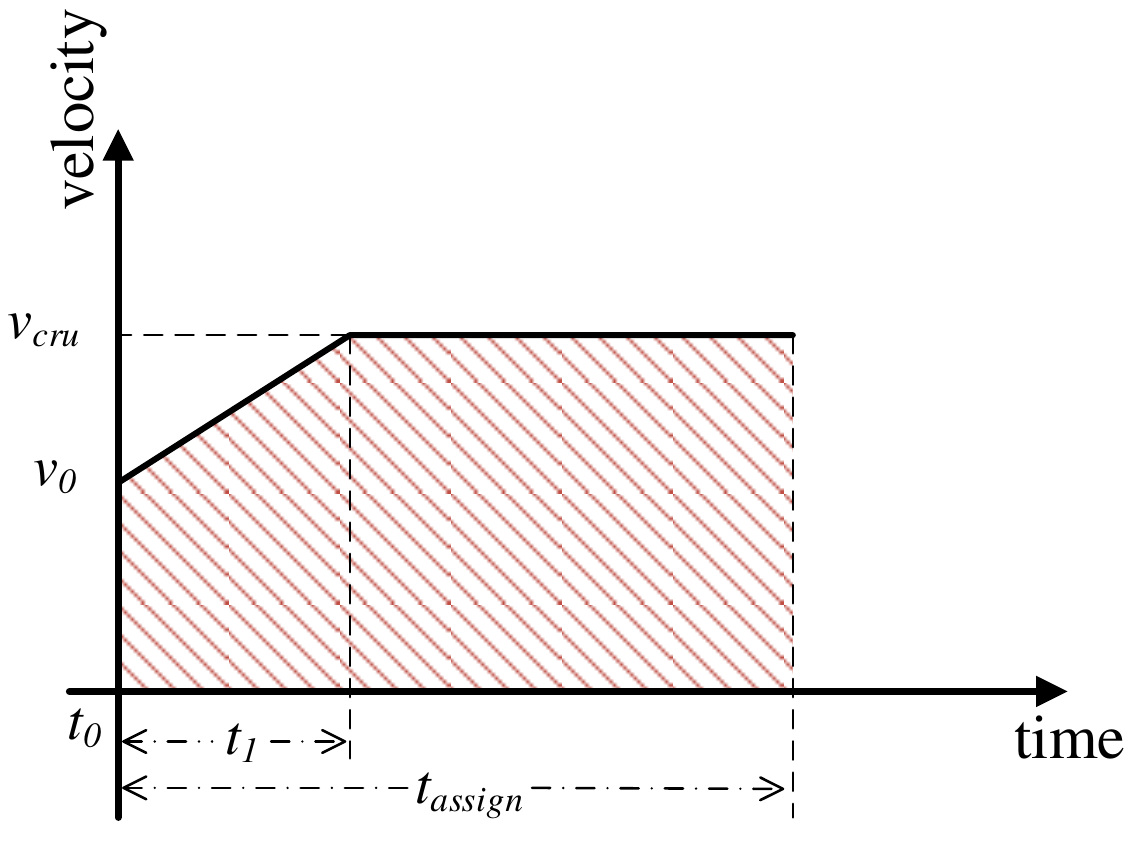}
  \end{minipage}
\end{lrbox}

\newsavebox\unione
\begin{lrbox}{\unione}
\small
  \begin{minipage}{4cm}
    \begin{align*}
        &a=a_{max}\\
        &t_1=t_{assign}-\Delta t^1\\
        &v_{cru}=v_0+a\cdot t_1\\
    \end{align*}
  \end{minipage}
\end{lrbox}

\newsavebox\unitwof
\begin{lrbox}{\unitwof}
  \begin{minipage}{4cm}
    \includegraphics[width=4cm]{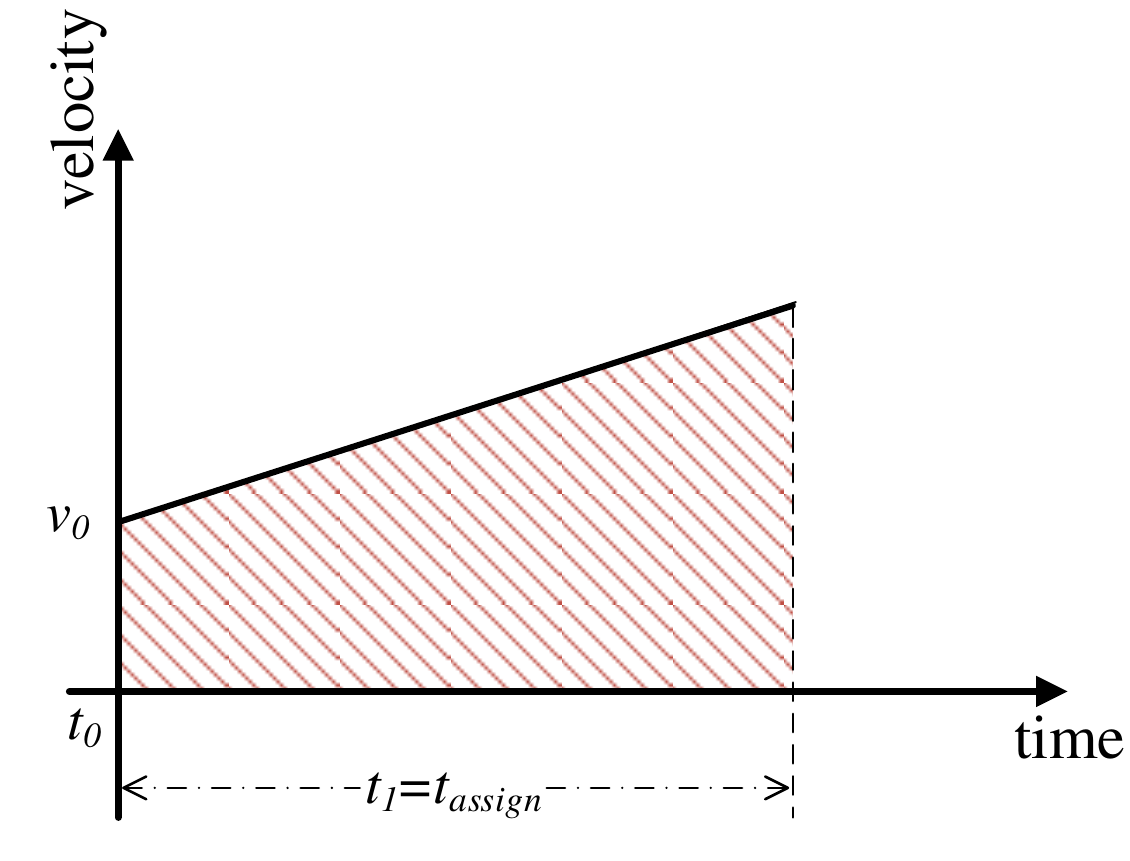}
  \end{minipage}
\end{lrbox}

\newsavebox\unitwo
\begin{lrbox}{\unitwo}
\small
  \begin{minipage}{4cm}
    \begin{align*}
        &a=\frac{2x_0-2v_0 \cdot t_{assign}}{t_{assign}^2}\\
        &t_1=t_{assign}\\
        &v_{cru}=v_0+a\cdot t_1\\
    \end{align*}
  \end{minipage}
\end{lrbox}

\newsavebox\unithreef
\begin{lrbox}{\unithreef}
  \begin{minipage}{4cm}
    \includegraphics[width=4cm]{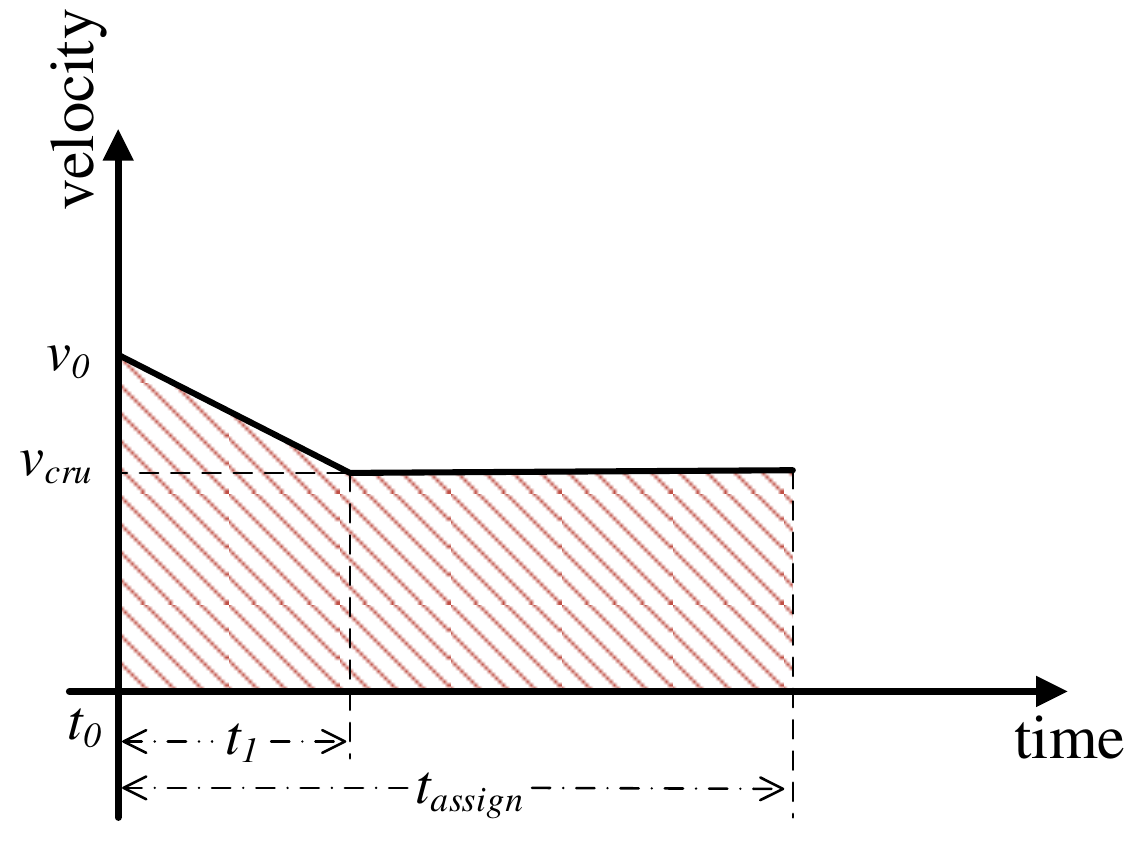}
  \end{minipage}
\end{lrbox}

\newsavebox\unithree
\begin{lrbox}{\unithree}
\small
  \begin{minipage}{4cm}
    \begin{align*}
        &a=a_{min}\\
        &t_1=t_{assign}+\Delta t^1\\
        &v_{cru}=v_0+a\cdot t_1\\
    \end{align*}
  \end{minipage}
\end{lrbox}

\newsavebox\unifourf
\begin{lrbox}{\unifourf}
  \begin{minipage}{4cm}
    \includegraphics[width=4cm]{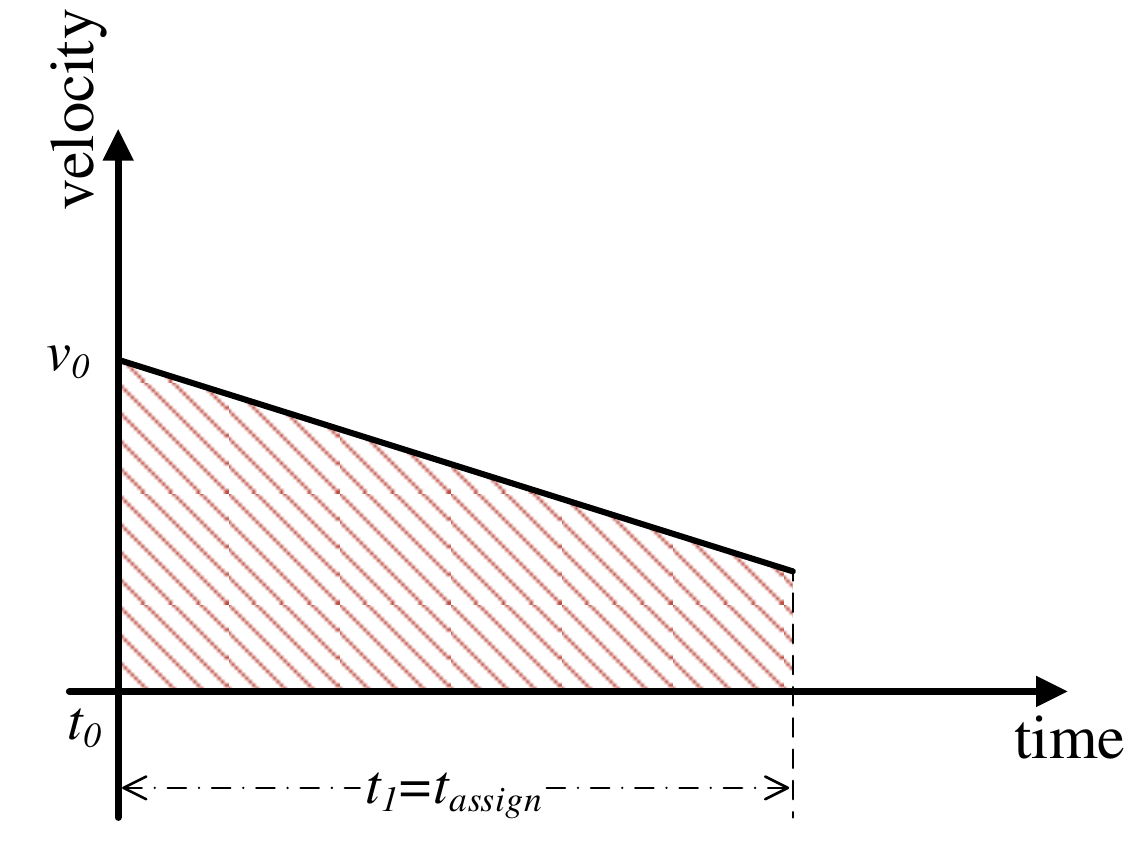}
  \end{minipage}
\end{lrbox}

\newsavebox\unifour
\begin{lrbox}{\unifour}
\small
  \begin{minipage}{4cm}
    \begin{align*}
        &a=\frac{2x_0-2v_0 \cdot t_{assign}}{t_{assign}^2}\\
        &t_1=t_{assign}\\
        &v_{cru}=v_0+a\cdot t_1\\
    \end{align*}
  \end{minipage}
\end{lrbox}

\newsavebox\unifivef
\begin{lrbox}{\unifivef}
  \begin{minipage}{4cm}
    \includegraphics[width=4cm]{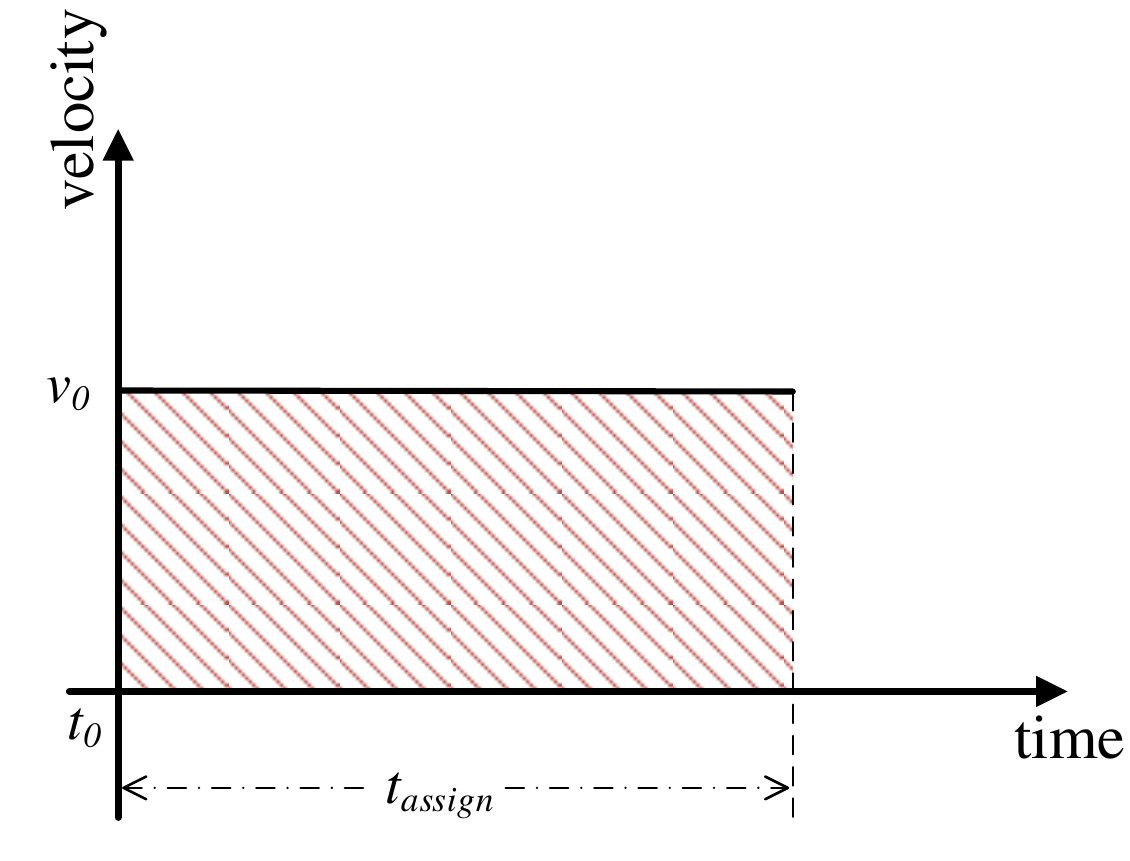}
  \end{minipage}
\end{lrbox}

\newsavebox\unifive
\begin{lrbox}{\unifive}
\small
  \begin{minipage}{4cm}
    \begin{align*}
        &a=0\\
        &t_1=0\\
        &v_{cru}=v_0\\
        &v_0\cdot t_{assign}=x_0\\
    \end{align*}
  \end{minipage}
\end{lrbox}

\begin{table}[H]
\footnotesize
  \centering
  %\caption{Motion planning strategy}
    \begin{tabular}{X{cc}|X{ll}}
    \toprule
    \multicolumn{2}{l}{Accelerating to a cruising velocity, then operating at the cruising velocity} \\
    \usebox{\unionef}     &  \usebox{\unione}   \\
    \midrule
    \multicolumn{2}{l}{Keeping accelerating.} \\
    \usebox{\unitwof}     &  \usebox{\unitwo}   \\
    \midrule
    \multicolumn{2}{l}{ Decelerating to a cruising velocity, then operating at the cruising velocity} \\
    \usebox{\unithreef}     &  \usebox{\unithree}   \\
    \midrule
    \multicolumn{2}{l}{Keeping decelerating} \\
    \usebox{\unifourf}     &  \usebox{\unifour}   \\
    \midrule
    \multicolumn{2}{l}{Keeping a constant velocity} \\
    \usebox{\unifivef}     &  \usebox{\unifive}   \\
    \bottomrule
    \end{tabular}%
    \begin{tablenotes}
            \footnotesize
            \item{$^1$} $\Delta t=\frac{\sqrt{(a\cdot t_{assign})^2+2a(v_0 \cdot t_{assign}-x_0)}}{a}$
    \end{tablenotes}
    \label{tab:motion}
\end{table}%

\bibliographystyle{IEEEtran}
\bibliography{IEEEabrv,reference}

\end{document}